\documentclass{article}
\usepackage[english]{babel}
\usepackage{amsfonts}
\usepackage{amssymb}
\usepackage[dvips]{graphicx}
\usepackage{subfigure}
\usepackage{color}

\begin{document}

\title{Finite sampling effects on generalized fluctuation-dissipation relations for steady states}
\author{Juan Ruben Gomez-Solano, Artyom Petrosyan and Sergio Ciliberto \\
Laboratoire de Physique, Ecole
Normale Sup\'erieure de  Lyon, CNRS UMR 5672,\\
        46, All\'ee d'Italie, 69364 Lyon CEDEX 07, France}
\maketitle


\begin{abstract}
We study the effects of the finite number of experimental data on the computation of a generalized fluctuation-dissipation relation around a nonequilibrium steady state of a Brownian particle in a toroidal optical trap. We show that the finite sampling has two different effects, which can give rise to a poor estimate of the linear response function. The first concerns the accessibility of the generalized fluctuation-dissipation relation due to the finite number of actual  perturbations imposed to the control parameter. The second concerns the propagation of the error made at the initial sampling of the external perturbation of the system.  This can be highly enhanced by introducing an estimator which corrects the error of the initial sampled condition. When these two effects are taken into account in the data analysis, the generalized fluctuation-dissipation relation is verified experimentally.

\end{abstract}

\section{Introduction}
Current theoretical developments in nonequilibrium statistical mechanics have led to significant progress in the study of systems around states far from thermal equilibrium. Systems in nonequilibrium steady states (NESS) are the simplest examples because the dynamics of their degrees of freedom $x$ under fixed control parameters $\lambda$ can be statistically described by time-independent probability densities $\rho_{0}(x,\lambda)$. NESS naturally occur in mesoscopic systems such as colloidal particles dragged by optical tweezeres, Brownian ratches and molecular motors because of the presence of nonconservative or time-dependent forces \cite{reimann}. At these lengthscales fluctuations are important so it is essential to establish a quantitative link between the statistical properties of the NESS fluctuations and the response of the system to external perturbations. Around thermal equilibrium this link is provided by the fluctuation-dissipation theorem \cite{marini}.

The generalization of the fluctuation-dissipation theorem around NESS for systems with Markovian dynamics has been achieved in recent years from different theoretical approaches \cite{harada,lippiello,seifert1,chetrite,prost,chetrite2,baiesi1,baiesi2,baiesi3,seifert2}. The different generalized formulations of the fluctuation-dissipation theorem link correlation functions of the fluctuations of the observable of interest $O(x)$ in the unperturbed NESS with the linear response function of $O(x)$ due to a small external time-dependent perturbation around the NESS. The observables involved in such relations are not unique but they are equivalent in the sense that they lead to the same values of the  linear response function. These theoretical relations may be  be useful in experiments and simulations to know the linear response of the system around NESS. Indeed the response can be obtaines from measurements entirely done at the unperturbed NESS of the system of interest without any need to perform the actual perturbation. Nevertheless, the theoretical equivalence of the different observables involved in those relations does not translate into equivalent experimental accessibility: \emph{e.g.} strongly fluctuating observables such as instantaneous velocities may lead to large statistical errors in the measurements \cite{mehl}. Besides, NESS quantities themselves such as local mean velocities, joint stationary densities and the stochastic entropy are not in general as easily measurable as dynamical observables directly related to the degrees of freedom \cite{gomez1}.  Hence, before implementing the different fluctuation-response formulae in real situations it is important to test its experimental validity under very well controlled conditions and to assess the influence of finite data analysis. The experimental test of some fluctuation-dissipation relations has been recently done in Refs.~\cite{mehl,gomez1,blickle1,gomez2} for colloidal particles in toroidal optical traps.

In the present paper we discuss the effects of the finite number of experimental data on the determination of the linear response function around a NESS for a micron-sized system with Markovian dynamics: a Brownian particle in a toroidal optical trap. For this purpose we perform the respective data analysis on the measurements reported in~\cite{gomez1,gomez2}. In Sect.~\ref{sec:HS} we briefly describe a generalized fluctuation-dissipation relation that has been derived for Markovian dynamics around a NESS exploiting the properties of the stationary density $\rho_0(x,\lambda)$. In Sect.~\ref{sec:particle} we recall the main features of a previous experiment that we use in the present paper for the data analysis. In Sect.~\ref{sec:GFD} we discuss the two different kinds of finite-sampling effects that can appear in the computation of the differents terms involved in the fluctuation-dissipation relation. We show that the generalized fluctuation-dissipation relation is verified experimentally when performing a careful data analysis, which takes into account these effects. Finally we present the conclusion.

\section{Hatano-Sasa relation and fluctuation-dissipation around NESS}\label{sec:HS}

The Hatano-Sasa relation provides a general identity for the transitions between either equilibrium or nonequilibrium steady states of Markovian systems \cite{hatano}. In the following we will focus on a Langevin system with Markovian dynamics described by a steady state probability density $\rho_0(x,\lambda)$. When the system is subjected to a time-dependent variation  of the control parameter $\lambda(t)$ between an initial time $t_i$ and a final time $t_f$, the Hatano-Sasa identity reads
\begin{equation}\label{eq:HatanoSasa}
    \left\langle \exp\left( -\int_{t_i}^{t_f} \mathrm{d}t \dot{\lambda}_{\alpha} (t) \frac{\partial \phi(x_t,\lambda(t))}{\partial \lambda_{\alpha}} \right)  \right\rangle = 1,
\end{equation}
where $\phi(x,\lambda)=-\ln \rho_0(x,\lambda)$ and the average $\langle \ldots \rangle$ is performed over an infinite number of realizations of a prescribed time-dependent protocol $\lambda(t)$. From Eq.~(\ref{eq:HatanoSasa}), Prost \emph{et al.} has directly derived a generalized fluctuation-dissipation relation that holds in the linear response regime around a NESS \cite{prost}
\begin{equation}\label{eq:GFDTJPP}
    R_{\alpha \gamma}(t-s) = \frac{d}{dt}\left\langle \frac{\partial \phi(x_t,\lambda_{SS})}{\partial \lambda_{\alpha}}\frac{\partial \phi(x_s,\lambda_{SS})}{\partial \lambda_{\gamma}} \right\rangle_0.
\end{equation}
In Eq.~(\ref{eq:GFDTJPP}) $R_{\alpha \gamma}(t-s)=\delta \langle O(x_t) \rangle_h / \delta h_s |_{h=0}$ is the linear response function of the observable $O(x)=\partial \phi(x,\lambda_{SS}) / \partial \lambda_{\alpha}$ due to a small external time-dependent perturbation $h_s=\lambda(s)-\lambda_{SS}$ around $\lambda_{SS}=\lambda(t_i)$ fixing an initial NESS at time $t_i$. The averages $\langle \ldots \rangle_{h}$ and $\langle \ldots \rangle_0$ are performed over the perturbed and unperturbed processes, respectively.

In experiments the formal average involved in Eq.~(\ref{eq:HatanoSasa}) is not perfectly computed because of the finite number of independent realizations of $\lambda(t)$. Hence Eq.~(\ref{eq:HatanoSasa}) allows one to estimate the experimental precision of (\ref{eq:GFDTJPP}) computed from a given number of experimental data provided that one can measure the observable $\partial \phi(x,\lambda_{SS}) / \partial \lambda_{\alpha}$. In the next section we tackle this problem for the experimental trajectories of a colloidal particle in a toroidal optical trap.

\begin{figure}
\centering
{\includegraphics[width=.375\textwidth]{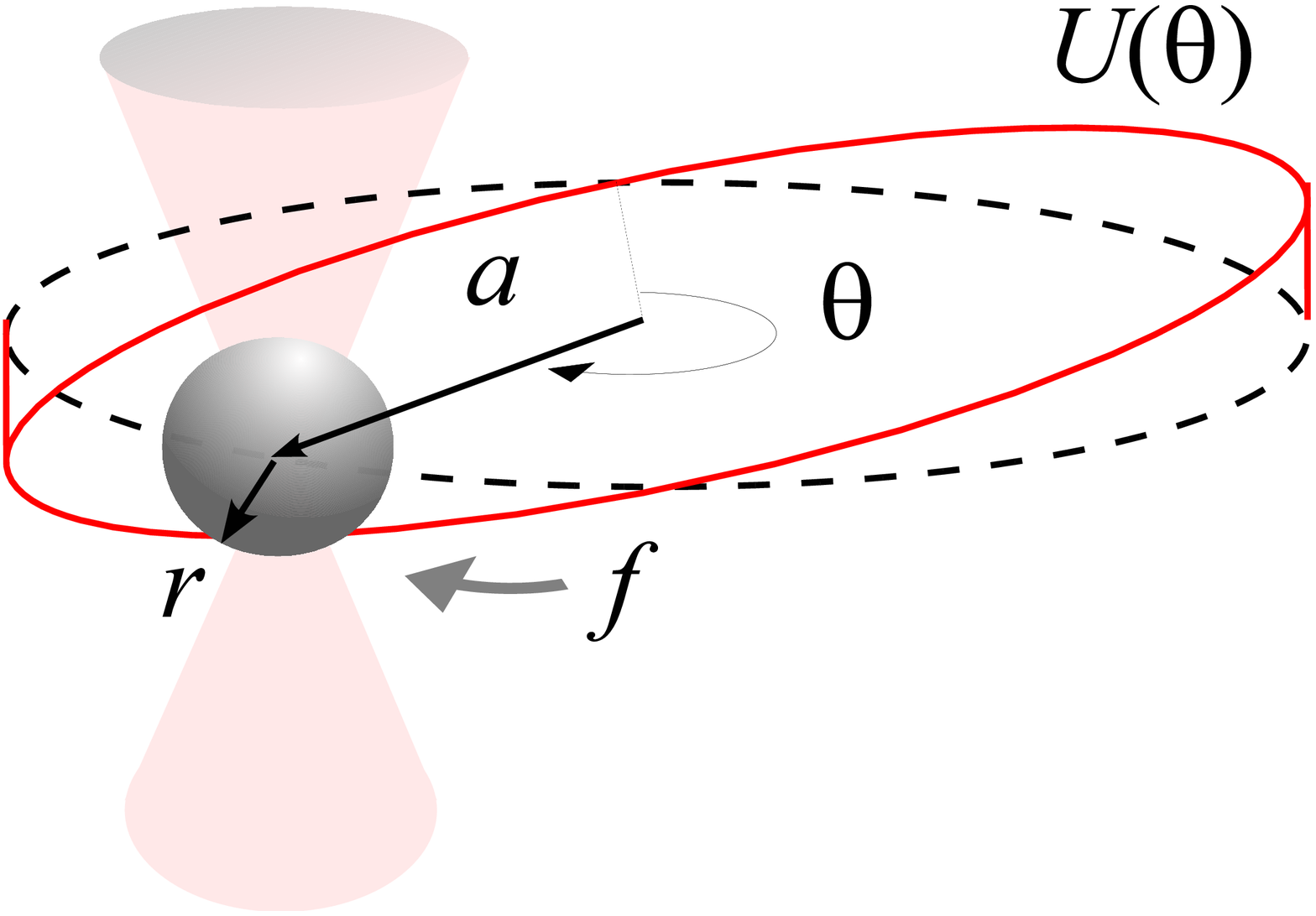}}
\hspace{.0in}
{\includegraphics[width=.375\textwidth]{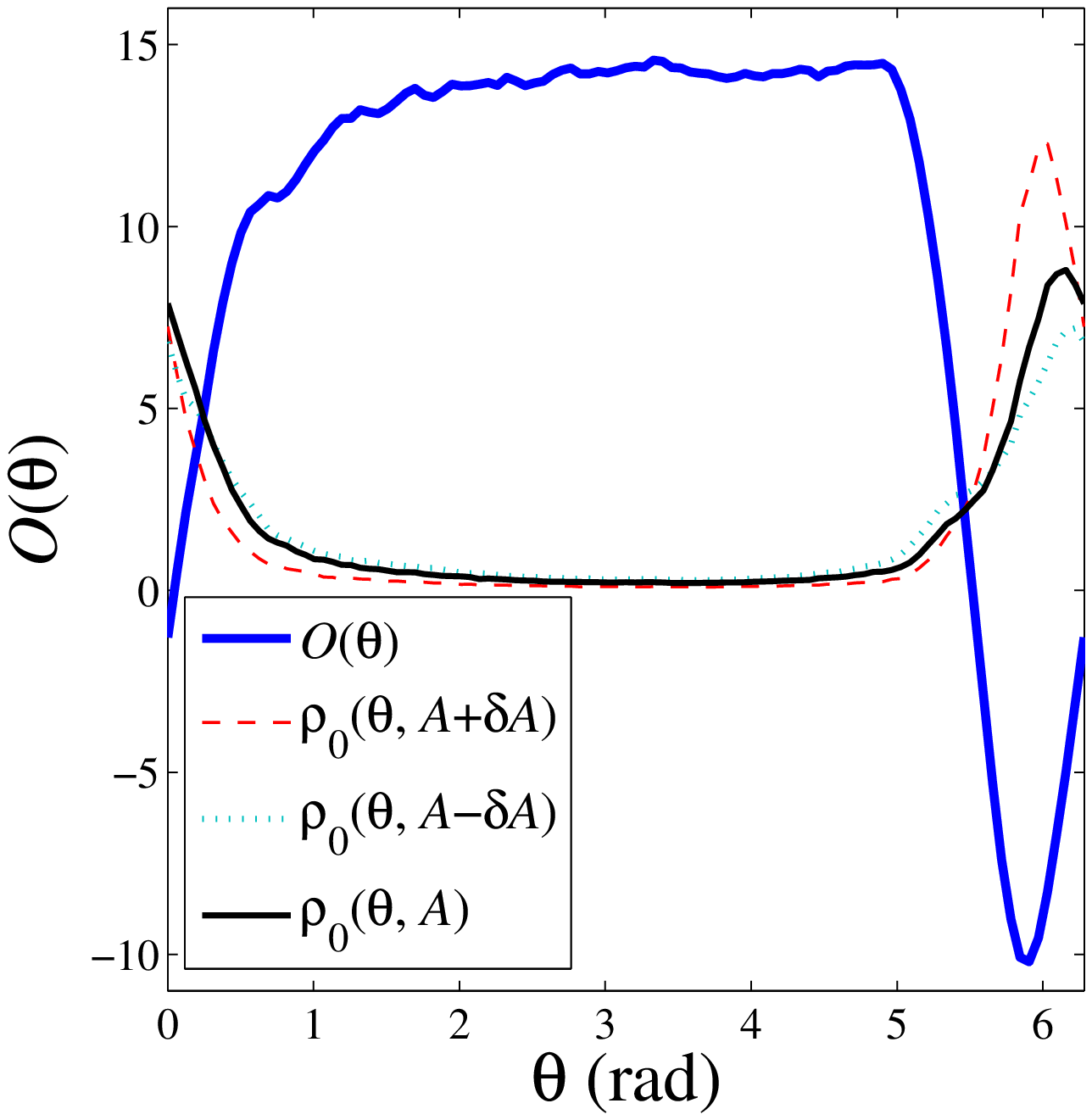}}
\hspace{.0in}
     \caption{(a) Sketch of a Brownian particle in a toroidal optical trap subjected to a nonconservative force $f$ and undergoing a periodic potential $U(\theta). $(b) Experimental profile of the observable $O(\theta)$ defined in Eq.~(\ref{eq:obs}) computed using the NESS densities $\rho_0(\theta,A+\delta A)$ and $\rho_0(\theta,A-\delta A)$ around $\rho_0(\theta,A)$ at fixed $F$.}
\label{fig:1}
\end{figure}

\section{Brownian particle in a toroidal optical trap}\label{sec:particle}
The Brownian motion of a colloidal particle in a toroidal optical
trap has become an experimental model to study the generalization
of the fluctuation-dissipation theorem around a
NESS~\cite{mehl,gomez1,blickle1,gomez2}. This is because it is a
system with a single translational degree of freedom where one can
easily tune its relevant control parameters. Our experiment has
been already described in detail in~\cite{gomez1,gomez2} so here
we only explain it briefly. The Brownian motion of a spherical
silica particle (radius $r=1 \,\mu$m) immersed in water is
confined on a thin torus of major radius $a=4.12 \, \mu$m by a
tightly focused laser beam rotating at 200 Hz. The water reservoir
acts as a thermal bath at fixed temperature ($T=20 \pm
0.5^{\circ}$C) providing thermal fluctuations to the particle. The
viscous drag coefficient at this temperature is $\gamma=1.89
\times 10^{-8}$ kg s$^{-1}$. The rotation frequency of the laser
is so high that it is not able to trap continuously the particle
in the focus because the viscous drag force of the surrounding
water quickly exceeds the optical trapping force. Consequently, at
each rotation the beam only kicks the particle a small distance
along the circle of radius $a$ exerting a nonconservative force
$f=66$ fN on it in the direction of the rotation. Thus, the
particle motion is effectively confined on a circle: the angular
position $\theta$ of its barycenter is the only relevant degree of
freedom. In addition, a static light intensity profile (amplitude
about $5\%$ of the total laser intensity~\cite{gomez1,gomez2}) is
created along the circle acting as a periodic potential
$U(\theta)=U(\theta+2\pi)$ of amplitude $68.8k_B T$.
Figure~\ref{fig:1}(a) depicts this experimental configuration. We
track the 2D particle position by video microscopy in order to
measure the time evolution of $\theta$. Thus for the
experimentally accessible length and time scales the dynamics of
$\theta_t$ is modeled by the first-order Langevin equation
\cite{mehl,gomez1,blickle1,gomez2,blickle2}
\begin{equation}\label{eq:1stLangevin}
\dot{\theta} = -\partial_{\theta}H(\theta)+F+ \xi,
\end{equation}
where $H(\theta) = U(\theta)/(\gamma a)$ with amplitude $A=\max\{H(\theta)\}=68.8k_BT/(\gamma a^2)$, $F = f / (\gamma a)$ and $\xi$ is a white noise process of zero mean and covariance
$\langle \xi_t \xi_s \rangle = 2[k_B T /(\gamma a^2)] \delta(t-s)$. Under these fixed conditions the dynamics of $\theta_t$ settles into a NESS whose probability density function $\rho_0(\theta,A)$ is plotted in Fig~\ref{fig:1}(b) (solid black line).

\section{Generalized fluctuation-dissipation relation}\label{sec:GFD}
In the following analysis we take $x=\theta$ as the single degree of freedom and $\lambda = A$ as the main control parameter of the system. In this case the observable of interest involved in the fluctuation-dissipation relation (\ref{eq:GFDTJPP}) is
\begin{equation}\label{eq:obs}
O(\theta)=-\frac{\partial \ln \rho_0(\theta,A)}{\partial A}.
\end{equation}
The experimental profile of $O(\theta)$, computed as a discrete three-point derivative of $-\ln \rho_0(\theta,A)$ at two different NESS around $A$, is shown in Fig~\ref{fig:1}(b).

We focus on the response of the system after applying a small Heaviside perturbation $\delta A$ to $A$ at time $t_i$: $A\rightarrow A + \delta A$. In the experiment this dynamical procedure is done by suddenly switching the laser power modulation as explained in \cite{gomez1,gomez2}. This procedure yields the integrated response function, defined as
\begin{equation}\label{eq:intresponse}
\chi(t-t_i)=\int_{t_i}^t R(t-s) \, \mathrm{d}s=\frac{\langle O(\theta_t) \rangle_{\delta A}}{\delta A},
\end{equation}
where the average $\langle \ldots \rangle_{\delta A}$ must be performed over the perturbed process at time $t$. Then the integrated version of the generalized fluctuation-dissipation relation~(\ref{eq:GFDTJPP}) is in this case
\begin{equation}\label{eq:intGFDT}
\chi(t-t_i) = \langle O(\theta_t) O(\theta_{t_i}) \rangle_0 - \langle O(\theta_{t}) O(\theta_t) \rangle_0.
\end{equation}
We now study the effects of a finite number of realizations of the perturbation $\delta A$ and the finite number of trajectories used to compute the averages $\langle \ldots \rangle_{\delta A}$ and $\langle \ldots \rangle_0$ in Eqs.~(\ref{eq:intresponse}) and (\ref{eq:intGFDT}) .

\subsection{Statistical error}
\begin{figure}
\centering
{\includegraphics[width=.45\textwidth]{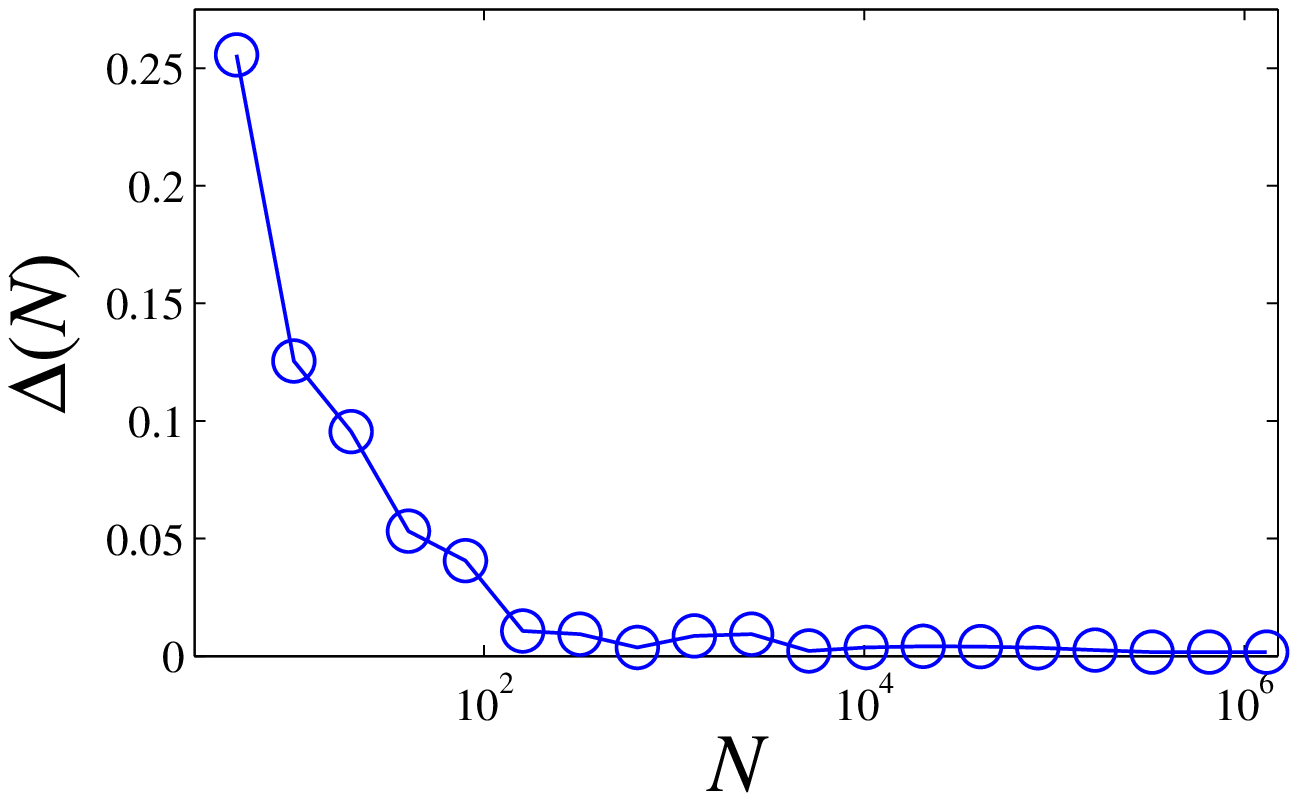}}
\hspace{.0in}
{\includegraphics[width=.45\textwidth]{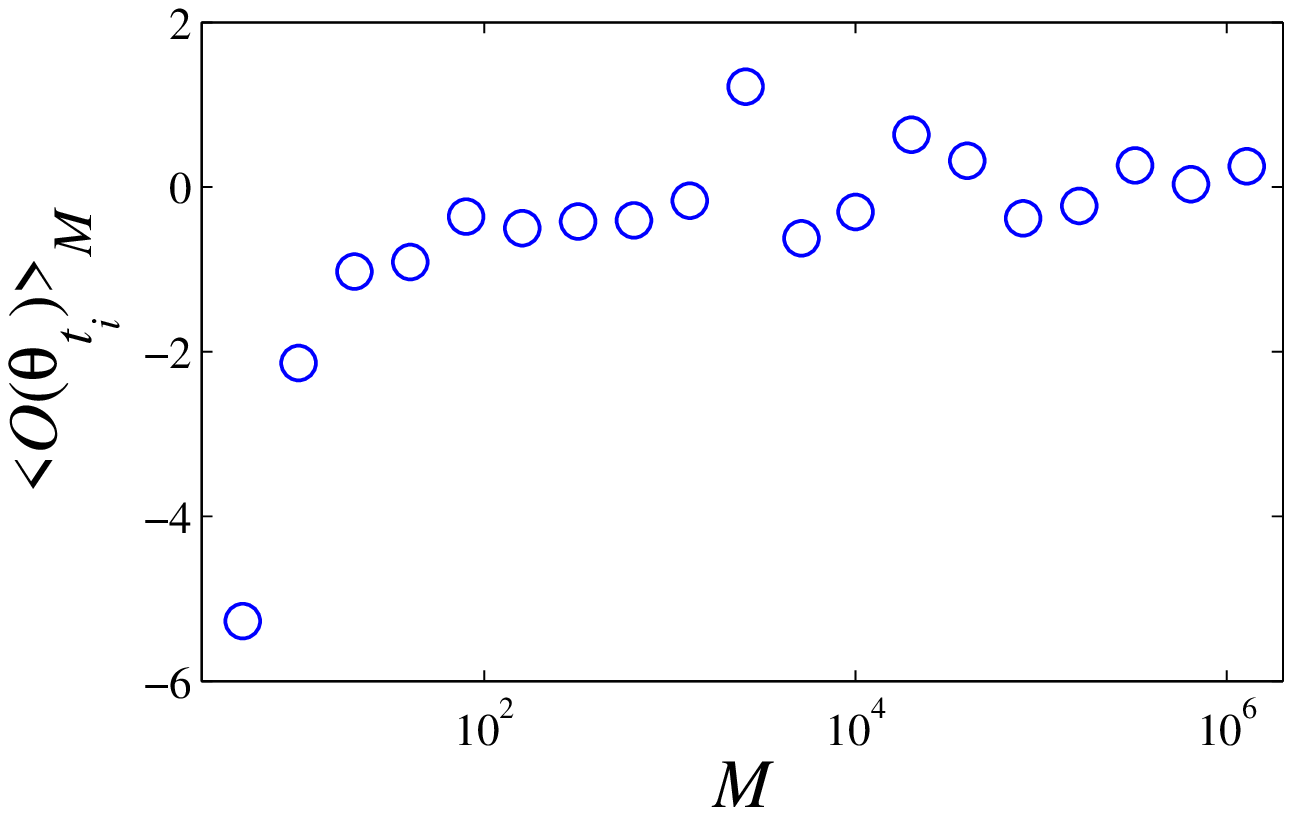}}\\
\vspace{.0in}
     \caption{(a) Estimate of the error of the Hatano-Sasa relation (\ref{eq:HS}) for a finite number $N$ of realizations of $\delta A$. (b) Estimate of the error of $\langle O(\theta_{t_i}) \rangle_0$ using Eq.~(\ref{eq:finiteinitresponse}) for $M$ NESS data.}
\label{fig:2}
\end{figure}

As discussed in Sect.~\ref{sec:HS}, the Hatano-Sasa relation (\ref{eq:HatanoSasa}) can be used to estimate the error of the experimental computation of Eq.~(\ref{eq:intGFDT}) when performing $N<\infty$ independent realizations of $\delta A$ around the NESS.
In the case of the dynamical process defined by the Heaviside perturbation $A \rightarrow A+  \delta A$ at time $t_i$, as done in the experiment, Eq.~(\ref{eq:HatanoSasa}) reads
\begin{equation}\label{eq:HS}
\left\langle \exp\left[ -\delta A O(\theta_{t_i})  \right] \right\rangle= 1.
\end{equation}
Eq.~(\ref{eq:HS}) only depends on the initial values $O(\theta_{t_i})$ when the system is still in NESS. Therefore, for a finite number of trajectories $N$ we introduce an estimator of the error of Eq.~(\ref{eq:HS})
\begin{equation}\label{eq:errorHS}
\Delta(N) = \left| \frac{1}{N}\sum_{j=1}^N \exp\left[ -\delta A O_j(\theta_{t_i})  \right] - 1\right|,
\end{equation}
where $O_j(\theta_{t_i})$ is the $j$-th sampled NESS initial condition. Fig.~\ref{fig:2}(a) shows the behavior of the error $\Delta(N)$ computed for $\delta A = 0.05 A$ (the value realized in the dynamical experiment) using $N$ experimental values of $O(\theta)$ drawn from the NESS distribution. For small $N \lesssim$ 100 the error is non-negliglible, $\Delta(N) \ge 4$\%, and this must be taken into account in the final accuracy of the generalized fluctuation-dissipation relation when comparing the left with the right-hand side of Eq.~(\ref{eq:intGFDT}) using the experimenal data. Then as $N$ increases $\Delta(N)$ quickly converges to 0: for $N \ge 500$ the precision of Eq.~(\ref{eq:HS}) found in the experiment is better than 1\%.

\subsection{Effect of the initial sampled condition}
According to the Heaviside procedure for $\delta A$, in Eqs.~(\ref{eq:intresponse}) and (\ref{eq:intGFDT}) the initial condition $\theta(t_i)$ for the perturbed process is sampled from the NESS density $\rho_0(\theta,A)$. Then the integreated linear response function must formally satisfy the initial condition
\begin{equation}\label{eq:initresponse}
\chi(0)=\frac{\langle O(\theta_{t_i}) \rangle_0}{\delta A}=0,
\end{equation}
where the last equality is due to the normalization of $\rho_0(\theta_{t_i},A)$.
We are interested in the effect of a finite number $M < \infty$ of initial values $\theta(t_i)$ drawn from the inital NESS on the estimate of $\chi(t-t_i)$. It should be noted that in practice a small $M$ may significantly affect the computation of $\langle O(\theta_{t_i}) \rangle_0$ in Eq.~(\ref{eq:initresponse}) because most of the positive values of $O(\theta)$ lie in the region where $\rho_0(\theta,A)$ is rarely sampled, as shown in Fig.~\ref{fig:1}(b). In Fig.~\ref{fig:2}(b) we plot some values of the finite average
\begin{equation}\label{eq:finiteinitresponse}
\langle O(\theta_{t_i}) \rangle_M = \frac{1}{M} \sum_{j=1}^M O_j(\theta_{t_i}),
\end{equation}
where $O_j(\theta_{t_i})$ is the $j$-th initial condition at NESS. As expected, for small $M$, $\langle O(\theta_{t_i}) \rangle_M < 0$ due to the fact that one samples mostly the negative values around the maximum of $\rho_0(\theta,A)$. The convergence to the theoretical value $\langle O(\theta_{t_i}) \rangle_0 = 0$ is very slow: as $M$ increases  $\langle O(\theta_{t_i}) \rangle_M$ becomes very sensitive to $M$ and large positive values of $\langle O(\theta_{t_i}) \rangle_M$ can be obtained. The general trend is around $\langle O(\theta_{t_i}) \rangle_0 = 0$, though. Then even for large $M$ one must be careful with the computation of the integrated response function since a large initial error of $\chi(0)$ due to the use of the average $\langle \ldots \rangle_M$ may significantly propagate as $t$ increases.

\begin{figure}[htp]
\centering
{\includegraphics[width=.315\textwidth]{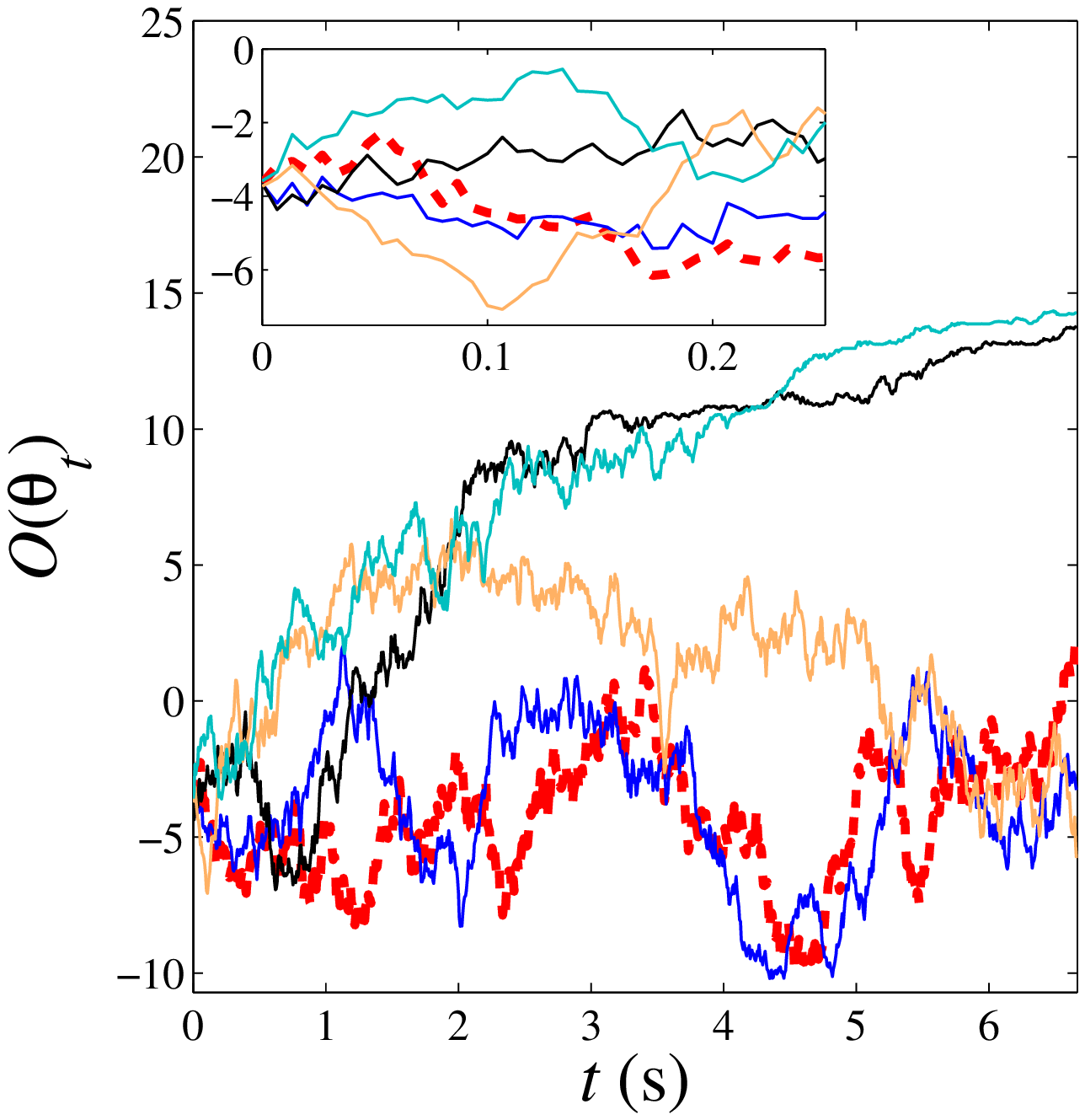}}
\hspace{.0in}
{\includegraphics[width=.315\textwidth]{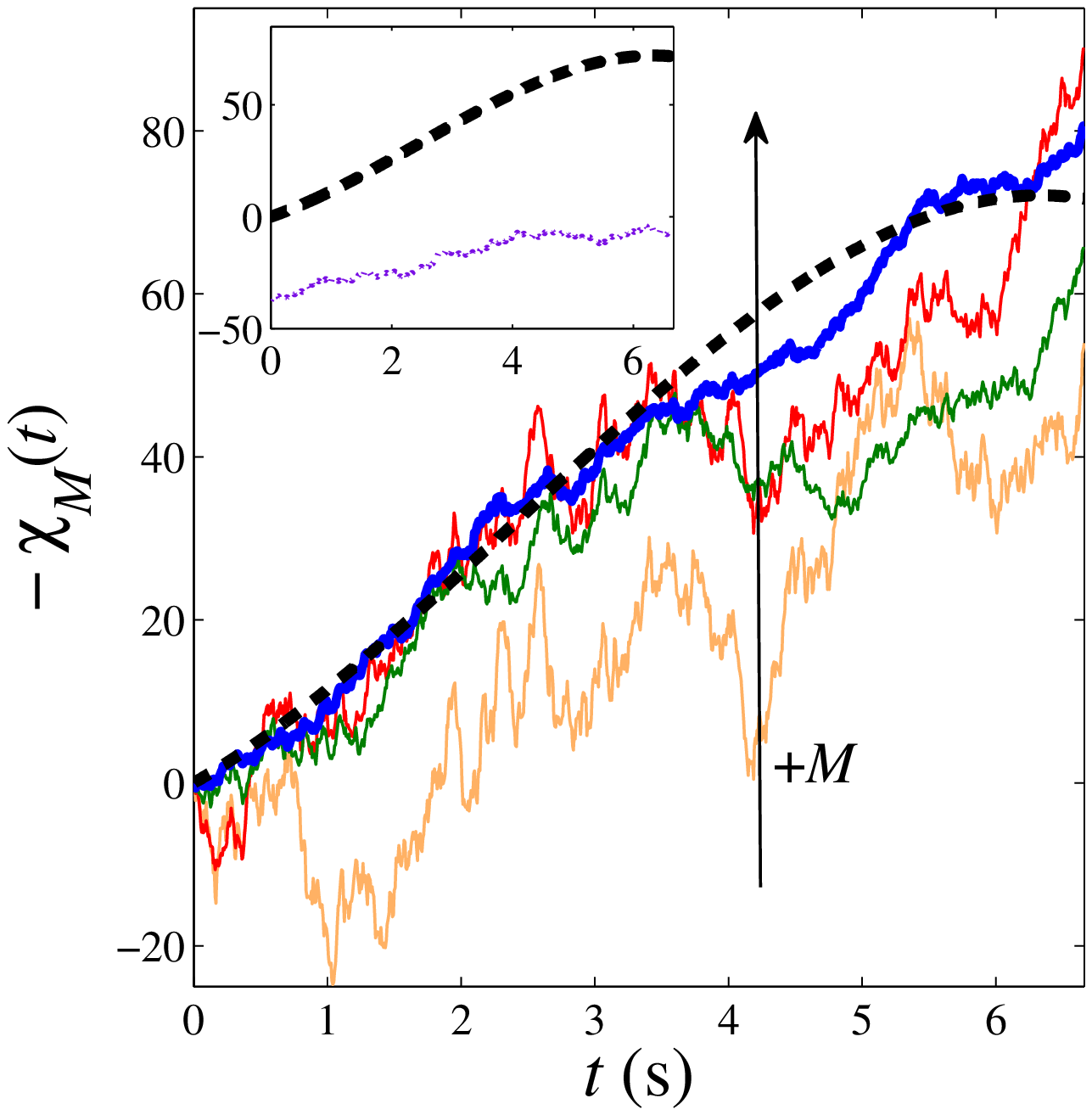}}
\hspace{.0in}
{\includegraphics[width=.315\textwidth]{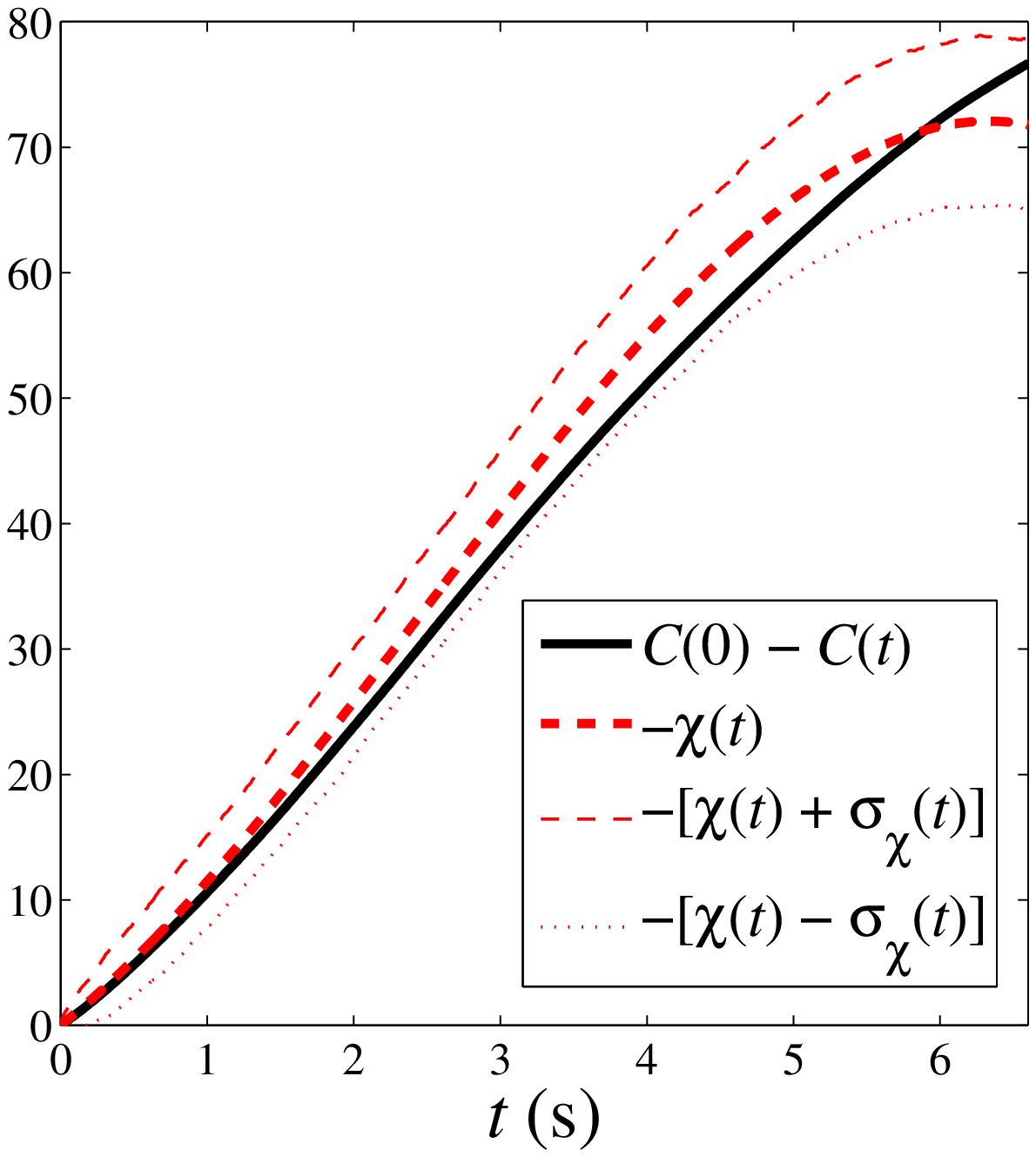}}
     \caption{(a) Examples of perturbed (thick dashed red line) and unperturbed (solid lines) trajectories used to estimate $\chi(t)$ using Eq.~(\ref{eq:estimatorresponse}). Inset: expanded view at short time. (b) Estimate of $-\chi(t)$ for $M=50,100,250,500$ and $L=1$ (solid
lines) and $M=500$, $L=200$ (dashed line). Inset: Comparison of $\chi_M(t)$ for $M=500$, $L = 200$ with the poor estimate done by the uncorrected average of Eq.~(\ref{eq:finiteinitresponse}) for $M=500$. (c) Comparison between the experimental $C(0)-C(t)$ and the best estimate of $-[\chi(t) \pm \sigma_{\chi}(t)]$.}
\label{fig:3}
\end{figure}

In order to avoid the problem of the  sensitivity to the initial
condition, instead of using directly the average $\langle \ldots
\rangle_M$ in Eq.~(\ref{eq:intresponse}), one can define an
estimator $\chi_M(t-t_{t_i})$ satisfying the initial condition
$\chi_M(0)=0$ as required ideally by Eq.~(\ref{eq:initresponse}).
In this way the propagation of the initial error given by $\langle
O(\theta_{t_i}) \rangle_M$  is suppressed at the beginning. An
intuitive way to define $\chi_M$ can be outlined from the usual
protocol to compute the integrated response function
\begin{equation}\label{eq:chi}
\chi(t-t_i)=\frac{\langle O^{\delta A}(\theta_{t}) \rangle_{\delta A}-\langle O(\theta_{(t-t_i+t^*)}) \rangle_{0}}{\delta A}
\end{equation}
where the time $t^*$ is chosen such that
$O(\theta_{t^*})=O^{\delta A}(\theta_{t_i})$ and   $O^{\delta A}(\theta_{t})$
denotes the observable measured during the perturbed process.
Notice that Eq.\ref{eq:chi} is justified by the fact that in the
case of an infinite number of samples $\langle O(\theta_{t})
\rangle_{0}=0 \ \forall t$, because of the time translational
invariance of the NESS.  In contrast when $M$ is finite, it is useful to take into account,in Eq.\ref{eq:chi},  that $\langle
\ldots \rangle_{\delta A}$ and $\langle \ldots \rangle_0$ are
performed independently the first on the perturbed trajectory
$O^{\delta A}(O(\theta_t)$ and the second on the unperturbed ones $O(\theta_{(t-t_i+t^*)})$, specifically
\begin{equation}\label{eq:estimator}
\chi_M(t-t_i) =\frac{1}{\delta A }\left[ \frac{1}{M} \sum_{j=1}^M O_j^{\delta A}(\theta_t) - \frac{1}{L}\sum_{k=1}^{L} O_{k}(\theta_{(t-t_i+t^*)})\right]
\end{equation}
where $L$ is the number of unperturbed trajectories such  $O_k(\theta_{t^*})=O_j^{\delta A}(\theta_{t_i})$.
Therefore Eq.\ref{eq:estimator} can be rewritten
\begin{equation}\label{eq:estimatorresponse}
\chi_M(t-t_i) =\frac{1}{\delta A }\frac{1}{M} \sum_{j=1}^M \left\{ \frac{1}{L}\sum_{k=1}^{L} \delta O_{jk}(\theta_t)\right\}
\end{equation}
where $\delta O_{jk}(\theta_t) \equiv O_j^{\delta A}(\theta_{t}) - O_k(\theta_{(t-t_i+t^*)})$ is the the instantaneous
difference between a perturbed trajectory $O_j^{\delta A}(\theta_t)$ and an
unperturbed  one $O_k(\theta_t)$.
An example of this procedure is depicted  in Fig.~\ref{fig:3}(a), where $t_i$ has been set equal to zero.  We see that for a given perturbed
trajectory $O_j^{\delta A}(\theta_t)$ (thick dashed red line),
obtained after that  $\delta A$ has been applied,
one should look for an unperturbed NESS
trajectory $O_k(\theta(t))$ such that $O_k(\theta_{t^*})=O_j^{\delta A}(\theta_{t_i})$
like the four unperturbed trajectories shown by the solid lines.

In this way  $\delta O_{jk}(\theta_{t_i})=0$ by construction and  the estimator defined by
Eq.~(\ref{eq:estimatorresponse}) satisfies the condition
$\chi_M(0)=0$. For $M,L \rightarrow \infty$,  $\chi_M(t-t_i)$  converges to
$\chi$ defined by Eq.~(\ref{eq:intresponse}),  because $\langle
\langle O(\theta_t)\rangle \rangle_M  \rightarrow 0$. In
Fig.~\ref{fig:3}(b) we show $\chi_M(t)$, with $t_i$ redefined as
$t_i=0$, computed using Eq.~(\ref{eq:estimatorresponse}) for
different values of $M$ and for fixed $L = 1$ (solid lines) and
$L=200$  (dashed line). As $M$ increases for $L = 1$ the
curves converge to a single profile which must correspond to that
of $\chi(t)$ ideally given by Eq.~(\ref{eq:intresponse}). The
additional conditional average done for $L=200$ smoothes the
slightly fluctuating profile for $M=500$ (thick solid blue line)
resulting in the thick dashed solid line. For comparison we also
show in the inset of Fig.~\ref{fig:3}(b) the raw estimate of
$\chi$ obtained using the average of
Eq.~(\ref{eq:finiteinitresponse}) for the same $M=500$ perturbed
trajectories without correcting the effect of the initial
sampling. In this case the propagation of the initial large error
of $\langle O(\theta_0) \rangle_0$ gives rise to a very poor
estimate of the integrated response function for $t>0$.

\subsection{Experimental test}
Finally we proceed to test the theoretical fluctuation-dissipation relation (\ref{eq:intGFDT}) for the experimental unperturbed NESS trajectories of the Brownian particle and those perturbed around NESS. For this purpose we compare the best estimate $\chi_M(t)$ of $\chi(t)$ done for $M=500$ and $L_j=200$ with the right-hand side of Eq.~(\ref{eq:intGFDT}). The involved correlation function $C(t) = \langle O(\theta_t) O(\theta_0) \rangle_0$ on the right-hand side is computed using unperturbed NESS trajectories. In Fig.~\ref{fig:3}(c) we compare $\chi_M(t)$ with $C(t)-C(0)$. Besides, one can estimate the statistical error of the experimental $\chi(t)$ at each $t \ge 0$ by computing the standard deviation of $\langle  O^{\delta A}(\theta_t) - O(\theta_t) \rangle_M$ over the $L_j$ possible choices of the unperturbed $O_k(t)$. The standard deviation $\pm \sigma_{\chi}(t)$ obtained in this way is also shown in Fig.~\ref{fig:3}(c) showing that after following the careful procedure to estimate $\chi$ the relation $\chi(t) = C(t)-C(0)$ is verified by the finite experimental data. Note that without the finite correction $\langle \langle O(\theta_t)\rangle\rangle_M$ of the initial condition in Eq.~(\ref{eq:estimatorresponse}) one would largely underestimate the direct measurement of the integrated response function leading to an apparent violation of Eq.~(\ref{eq:intGFDT}). The results of the present paper are consistent with those of Refs.~\cite{gomez1,gomez2} where two fluctuation-dissipation formulae equivalent to Eq.~(\ref{eq:intGFDT}) but involving different observables from the one studied here are checked experimentally.

\section{Conclusion}
We have studied the influence of finite sampling in the computation of the linear response function of a Brownian particle in a toroidal optical trap around a NESS. We have shown that there are two different effects that may lead to a very poor estimate of the experimental linear response function when the data analysis is not performed carefully. This is an important point that must be assessed in general when applying in experiment and numerical simulations the different generalized fluctuation-dissipation formulae  recently derived for  NESS.

\end{document}